\documentclass[a4paper,floatfix,rmp,twocolumn,showkeys,superscriptaddress]{revtex4}
  \usepackage[latin1]{inputenc}
  \usepackage{graphicx}

\begin{document}

  \title{Information theory, predictability, and the emergence of complex life}

  \providecommand{\MIT}{Department of Physics, Massachusetts Institute of Technology, Cambridge, MA 02139. }
  \providecommand{\ICREA}{ICREA-Complex Systems Lab, Universitat Pompeu Fabra (GRIB), Dr Aiguader 80, 08003 Barcelona, Spain. }
  \providecommand{\IBE}{Institut de Biologia Evolutiva, CSIC-UPF, Pg Maritim de la Barceloneta 37, 08003 Barcelona, Spain. }
  \providecommand{\SFI}{Santa Fe Institute, 1399 Hyde Park Road, Santa Fe NM 87501, USA. }
  
  \author{Lu\'is F Seoane}
    \affiliation{\MIT}
    \affiliation{\ICREA}
    \affiliation{\IBE}
  \author{Ricard V. Sol\'e}
    \affiliation{\ICREA}
    \affiliation{\IBE}
    \affiliation{\SFI}

  \vspace{0.4 cm}
  \begin{abstract}
    \vspace{0.2 cm}

    Despite the obvious advantage of simple life forms capable of fast replication, different levels of cognitive complexity have been achieved by living systems in terms of their potential to cope with environmental uncertainty. Against the inevitable cost associated to detecting environmental cues and responding to them in adaptive ways, we conjecture that the potential for predicting the environment can overcome the expenses associated to maintaining costly, complex structures. We present a minimal formal model grounded in information theory and selection, in which successive generations of agents are mapped into transmitters and receivers of a coded message. Our agents are guessing machines and their capacity to deal with environments of different complexity defines the conditions to sustain more complex agents. 

  \end{abstract}


\maketitle

 \section{Introduction}
    \label{sec:1}

    Simple life forms dominate our biosphere \cite{Gould2011} and define a lower bound of embodied, self-replicating systems. But life displays an enormously broad range of complexity levels, affecting many different traits of living entities, from their body size to their cognitive abilities \cite{Bonner1988}. This creates somewhat a paradox: if larger, more complex organisms are more costly to grow and maintain, why is not all life single-celled? Several arguments help provide a rationale for the emergence and persistence of complex life forms. As an instance \citet{Gould2011} proposes that complexity is not a trait explicitly favored by evolution. A review of fossil records convinces Gould that, across genera, phyla, and the whole biosphere, we observe the expected random fluctuations around the more successful adaptation to life. In this big picture, bacteria are the leading life form and the complexity of every other living system is the product of a random drift. Complex life would never be explicitly favored, but a complexity wall exists right below bacteria: simpler forms fail to subsist. Hence, a random fluctuation is more likely to produce more complex forms, falsely suggesting that evolution promotes complexity. 

    Major innovations in evolution involve the appearance of new types of agents displaying cooperation while limiting conflict \cite{MajorT1997,ReplicatorsReproducers}. A specially important innovation involved the rise of cognitive agents, namely those capable of sensing their environments and reacting to their changes in a highly adaptable way \cite{Jablonka 2006}. These agents were capable of dealing with more complex, non-genetic forms of information. The advantages of such cognitive complexity become clear when considering their potential to better predict the environment, thus reducing the average hazards of unexpected fluctuations. As pointed by Francois Jacob, an organism is ``a sort of machine for predicting the future -- an automatic forecasting apparatus'' \cite{Jacob1998, Wagensberg2000, Friston2013}. The main message is that {\em foreseeing} the future is a crucial ability to cope with uncertainty. If the advantages of prediction overcome the problem of maintaining and replicating the costly structures needed for inference, more complex information-processing mechanisms might be favored under the appropriate circumstances.

    \begin{figure*}
      \begin{center}
        \includegraphics[width=0.95 \textwidth]{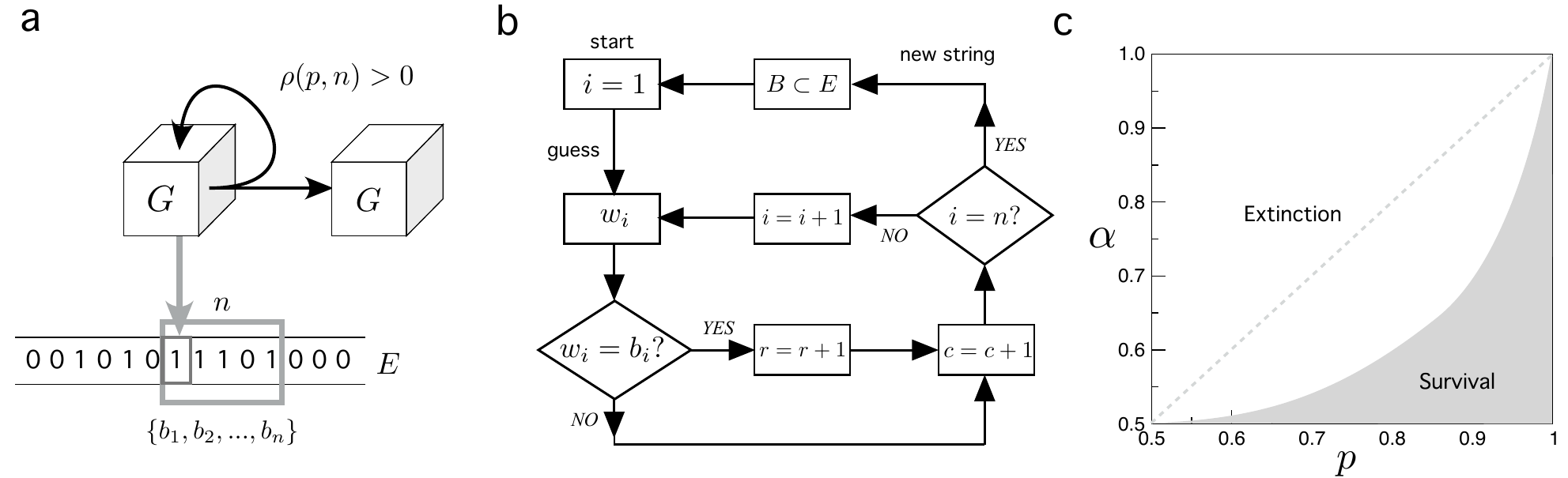}

        \caption{{\bf Predictive agents and environmental complexity.} {\bf a} An agent $G$ interacts with  an external environment $E$ that is modeled as a string of random bits. These bits take value $0$ with probability $p$ and value $1$ otherwise. The agent tries to guess a sequence of $n$ bits at some cost, with a reward bestowed for each correctly guessed bit. The persistence and replication of the agent can only be granted if the balance between reward and cost is positive ($\rho^G_E>0$). {\bf b} For a machine attempting to guess $n$ bits, an algorithmic description of its behavior is  shown as a flow graph. Each loop in the computation involves scanning a random subset of the  environment $B = (b_1, ..., b_n) \subset E$ by comparing each $b_i \in B$ to a proposed guess $w_i$.  {\bf c} A mean field approach to a certain kind of $1$-guesser (modeled in the text through equations \ref{eq:01}, \ref{eq:02}, and \ref{eq:03}) in environments of infinite size renders a boundary between survival ($\rho^G_E>0$) and death ($\rho^G_E<0$) as a function of the cost-reward ratio  ($\alpha$) and of relevant parameters for the $1$-guesser model ($p$ in this case). Note that for $\alpha < 0.5$ every $1$-guesser survives for free. }

        \label{fig:1}
      \end{center}
    \end{figure*}

    Here we aim at providing a minimal model that captures these tradeoffs. In doing so, we characterize thoroughly an evolutionary driver that can push towards evermore complex life forms. We adopt an information theory perspective in which agents are inference devices interacting with a Boolean environment. For convenience, this environment is represented by a tape with ones and zeros, akin to non-empty inputs of a Turing machine (figure 1{\bf a}). The agent $G$ locates itself in a given position and tries to predict each bit of a given sequence of length $n$ -- hence it is dubbed an {\em n-guesser}. Each attempt to predict a bit involves some cost $c$, while a reward $r$ is received for each successful prediction. $1$-guessers are simple and assume that all bits are uncorrelated, while $(n>1)$-guessers find correlations and can get a larger benefit if some structure happens to be present in the environment. A whole $n$-bit prediction cycle can be described as a program (figure 1{\bf b}). A survival function $\rho$ depends on the number of attempts to guess bits and the number of correct predictions. Successful guessers have a positive balance between reward and prediction cost. They get replicated and pass on their inference abilities. Otherwise, the agent fails to replicate and eventually dies.

    As a simple illustration of our approach, consider a $1$-guesser living in an infinitely large environment $E$ where uncorrelated bits take value $0$ with probability $p$ and $1$ with probability $1-p$. The average performance of a guesser $G$ when trying to infer bits from $E$ is given by $\bar{p}^G_E$, the likelihood of emitting a correct guess:
      \begin{eqnarray}
        \bar{p}^G_E &=& p^G(0)p + p^G(1)(1-p),
        \label{eq:01}
      \end{eqnarray}
    where $p^G(k)$ is the frequency with which the guesser emits the bit value $k \in \{0,1\}$. A strategy that uses $p^G(0)=p$, $p^G(1)=1-p$ (i.e. a guesser that mimics the environment) makes on average 
      \begin{eqnarray}
        \bar{p}^G_E &=& 2p^2-2p+1
        \label{eq:02}
      \end{eqnarray}
    successful predictions. Its survival function reads: 
      \begin{eqnarray}
        \rho^G_E = (2p^2-2p+1)r - c. 
        \label{eq:03}
      \end{eqnarray}
    This curve trivially dictates the average survival or extinction of $1$-guessers as a function of the cost-reward ratio $\alpha \equiv c/r$. Note that any more complex guesser (like the ones described below) would always fare worst in this case: they would potentially pay a larger cost to infer some structure where none is to be found. Note also that the tunable parameter $\alpha$ codes for the severity of the environment.

    The idea of autonomy and the fact that predicting the future implies performing some sort of computation suggests that a parsimonious theory of life's complexity needs to incorporate reproducing individuals (and eventually populations) and information (they must be capable of predicting future environmental states). These two components define a conflict and an evolutionary tradeoff. Being too simple means that the external world is perceived as a source of noise. Unexpected fluctuations can be harmful and useful structure cannot be harnessed in your benefit. Becoming more complex (hence able to infer larger structures, if they exist) implies a risk of not being able to gather enough energy to support and replicate the mechanisms for inference. As will be shown below, it is possible to derive the critical conditions to survive as a function of the agent's complexity and to connect these conditions to information theory. As advanced above, this allows us to characterize mathematically a scenario in which a guesser's complexity is explicitly selected for. 

  \section{Evolution and Information Theory}
    \label{sec:2}

    Key aspects of information theory relate deeply to formulations in statistical physics \cite{Jaynes1957a, Jaynes1957b, ParrondoSagawa2015} and there have been several calls to further integrate information theory in biological research \cite{MaynardSmith2000, Joyce2002, Nurse2008, Krakauer2011, WalkerDavies2012, Joyce2012}. This theory shall play important roles in population or ecosystems dynamics, in regulatory genomics, and in chemical signal processing among others \cite{McNamaraHouston1987, Szathmary1989, SegreLancet2000, SegreLancet2001, DallJohnstone2002, BergstromLachmann2004, DallStephens2005, KussellLeibler2005, DonalsonMatasciBergstrom2008, DonaldsonLachmann2010, RivoireLeibler2011, Adami2012, Friston2013, TkacikBialek2014, HidalgoMaritan2014, EvansDall2015, SartoriHorowitz2014, MarzenDeDeo2016}, but a unifying approach is far from complete.  Given its generality and power, information theory  has also been used to address problems that connect Darwinian evolution and far from equilibrium  thermodynamics \cite{NicolisPrigogine1977, Drossel2001, GoldenfeldWoese2010, England2013, PerunovEngland2014}. In its original formulation, Shannon's information theory \cite{Shannon1948, ShannonWeaver1949}  considers symbols being conveyed from a transmitter to a receiver through a channel. Shannon only deals with the efficiency of the channel (related to its noise or reliability) and the entropy of the source. This theory ignores the content of the emitted symbols, despite the limitations of such an assumption \cite{MaynardSmith2000, CorominasMurtraSole2014}.

\begin{figure}
      \begin{center}
        \includegraphics[width=0.35\textwidth]{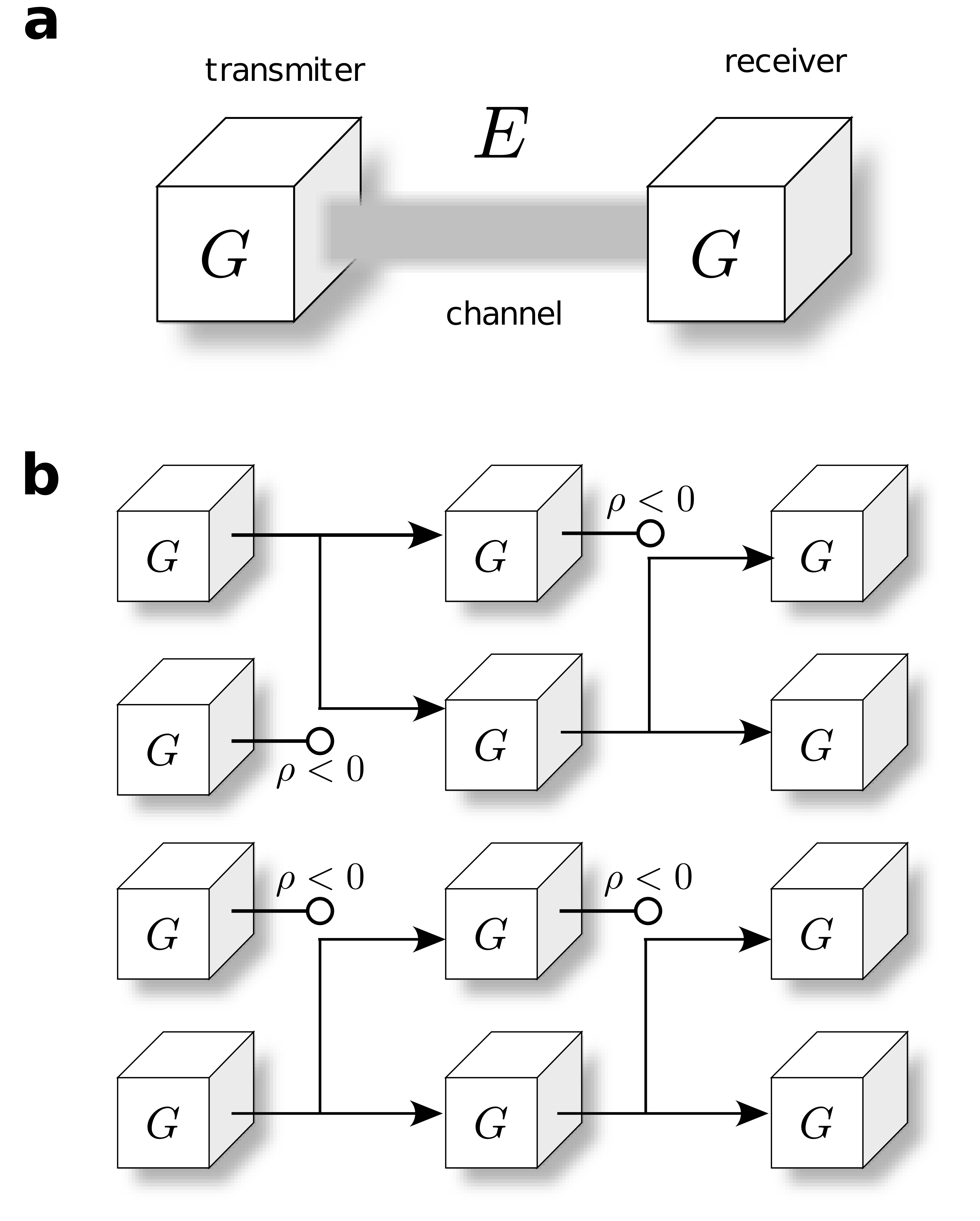}

        \caption{{\bf Information and evolution through natural selection.} {\bf a} The propagation of a successful replicator can be understood in terms of a Shannon-like transmission process from one generation to the next in which older generations play the role of a transmitter, younger generations that of a receiver, and the environment constitutes a noisy channel.  {\bf b} A simple diagram of the underlying evolution of a population of bit guessers. The survival and replication of a given agent $G$ is indicated by branching whereas failure to survive is indicated with an empty circle as and endpoint. }

        \label{fig:2}
      \end{center}
    \end{figure}

    A satisfactory connection between natural selection and information theory can be obtained by mapping our survival function $\rho$ into Shannon's transmitter-receiver scheme. To do so we consider replicators at an arbitrary generation $T$ attempting to ``send'' a message to (i.e. getting replicated into) a later generation $T+1$. Hence, the older generation acts as a transmitter, the newer one becomes a receiver, and the environment and its contingencies constitute the channel through which the embodied message must be conveyed (figure \ref{fig:2}{\bf a}). From a more biological perspective, we can think of a genotype as a generative model (the instructions in an algorithm) that produces a message that must be transmitted. That message would be embodied by a phenotype and it includes every physical process and structure dictated by the generative model. As discussed by \citet{vonNeumanBurks1966}, any replicating machine must pass on a physically embodied copy of its instructions -- hence the phenotype must also include a physical realization of the algorithm encoded by the genotype\footnote{Note that many of the phenotypic structures built in order to get replicated are later dismissed (think, e.g., about the germ vs. somatic cell lines). We present a clear division between genotype and phenotype for sake of illustration. We are aware of the murky frontier between these concepts. }. Finally, any evolutionary pressure (including the interaction with other replicating signals) can be included as contrivances of the channel.

    Following a similar idea of messages being passed from one generation to the next one, \citet{MaynardSmith2000} proposes that the replicated genetic message carries {\em meaningful information} that must be protected {\em against} the channel contingencies. Let us instead depart from a replicating message devoid of meaning. We realize that the channel itself would convey more reliably those messages embodied by a phenotype that better deals with the environmental (i.e. channel) conditions. Dysfunctional messages are removed due to natural selection. Efficient signals get more space in successive generations (figure \ref{fig:2}{\bf b}). Through this process {\em meaningful} bits of environmental information are {\em pumped} into the replicating signals, such that the information in future messages will {\em anticipate} those channel contingencies. In our picture, meaningful information is not protected against the channel conditions (including noise), but emerges naturally from them.

    \subsection{Messages, channels, and bit guessers}
      \label{sec:2.1}

\begin{figure}
        \begin{center}
          \includegraphics[width=0.45\textwidth]{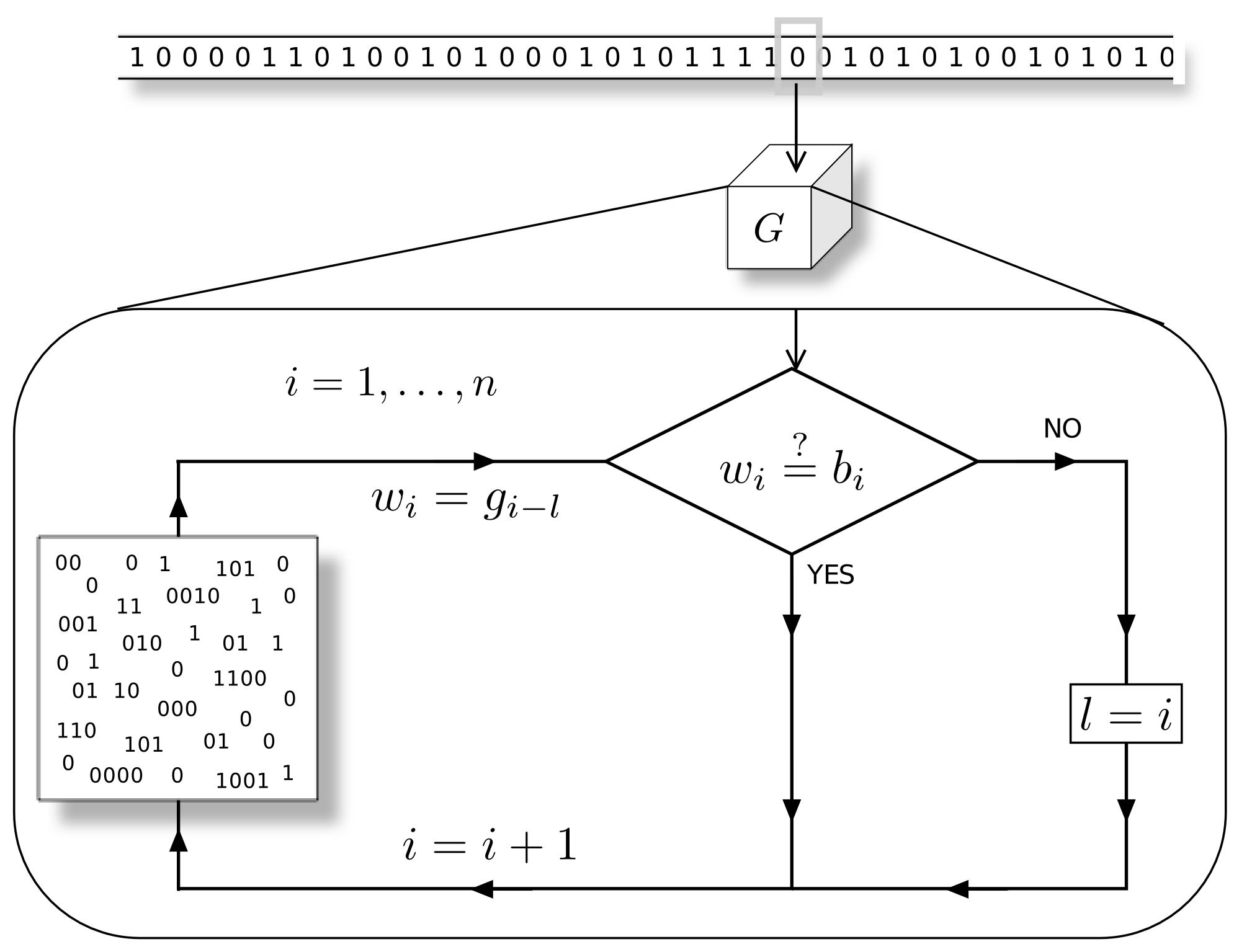}
          \caption{{\bf From a generative model to inference about the world. } A diagrammatic representation of the algorithmic logic of the bit guessing machine. Our $n$-guesser contains a generative model (represented by a pool of words) from which it draws guesses about the environment. If a bit is successfully inferred, the chosen conjecture is pursued further by comparing a new bit. Otherwise, the inference is reset.}

          \label{fig:3}
        \end{center}
      \end{figure}

      Let us first introduce our implementation of environments (channels), messages, and the replicating agents. The later will be dubbed {\em bit-guessers} because efficient transmission will be equivalent to accurately predicting channel conditions -- i.e. to correctly guessing as many bits about the environment as possible. The notation that follows may seem arid, so it is good to retain a central picture (figure \ref{fig:3}): Guessers $G$ posses a generative model $\Gamma^G$ that must produce messages that fare well in an environment $E$. Both these messages and the environments are modeled as strings of bits. What follows is a rigorous mathematical characterization of how the different bit sequences are produced. 

      Let us consider $m$-environments, strings made up of $m$ sorted random bits. We might consider one single such $m$-environment -- i.e. one realization $E$ of $m$ sorted random bits ($e_i \in E$, $i=1, \dots, m$; $e_i \in \{0,1\}$). Alternatively, we might work with the ensemble $E^m$ of all $m$-environments -- i.e. all possible environments of the same size ($e_{i,l} \in E_l$, $i=1, \dots, m$; where $E_l \in E^m$, $l=1, \dots, 2^m$) -- or we might work with a sample $\hat{E}^m$ of this ensemble ($E_l \in \hat{E}^m$, $l = 1, \dots, ||\hat{E}^m||$; where $\hat{E}^m \subset E^m$). We might evaluate the performance of our bit guessers in single $m$-environments, in a whole ensemble, or in a sample of it. 

      These $m$-environments model the channels of our information theory approach. Attempting to transmit a message through this channel will be implemented by trying to guess $n$-sized words from within the corresponding $m$-environment. More precisely, given an $n$-bit message $W$ (with $n<m$) which an agent tries to transmit, we extract an $n$-sized word ($B \subset E$) from the corresponding $m$-environment. Therefore, we choose a bit at a random position in $E$ and the successive $n-1$ bits. These make up the $b_i \in B$, which are compared to the $w_i \in W$. Each $w_i$ is successfully transmitted through the channel if $w_i = b_i$. Hence attempting to transmit messages effectively becomes an inference task: if a guesser can anticipate the bits that follow, it has a greater chance of sending messages through. Messages transmitted equal bits copied into a later generation, hence increasing the fitness of the agent. \\

      In this paper we allow bit-guessers a minimal ability to react to the environment. Hence, instead of attempting to transmit  a fixed word $W$, they are endowed with a generative model $\Gamma^G$. This mechanism (explained below) builds the message $W$ as a function of the broadcast history: $$w_i = w_i(w_1, \dots, w_{i-1}; b_1, \dots, b_{i-1}).$$ Hence, the fitness of a generative model is rather based on the ensemble of messages that it can produce\footnote{There is a compromise worth investigating between the fidelity of the message that an agent tries to convey and its ability to react to environmental conditions in real time. Exploring this tradeoff is left for future work. By now, the reaction capabilities of our bit-guessers will be kept to a minimum.}. To evaluate this, our guessers attempt to transmit $n$-bit words many ($N_g$) times through a same channel. For each one of these broadcasts, a new $n$-sized word $B^j \subset E$ (with $b^j_i \in B^j$ for $j = 1, \dots, N_g$ and $i=1, \dots, n$) is extracted from the same $m$-environment; and the corresponding $W^j$ are generated, each based on the broadcast history as dictated by the generative model (see below). 

      We can calculate different frequencies with which the guessers or the environments present bits with value $k,k' \in \{0,1\}$:
        \begin{eqnarray}
          \label{eq:1}
          p^{G}(k; i) &=& {1\over N_g} \sum_{j=1}^{N_g} \delta (w^j_i, k), \\
          \label{eq:2}
          p_E(k'; i) &=& {1\over N_g} \sum_{j=1}^{N_g} \delta(b^j_i, k'), \\
          \label{eq:3}
          p_{G,E}(k, k'; i) &=& {1\over N_g} \sum_{j=1}^{N_g} \delta(w^j_i, k)\delta(b^j_i, k'), \\
          \label{eq:4}
          p^G_E(i) &=& {1\over N_g} \sum_{j=1}^{N_g} \delta(w^j_i, b^j_i) \Rightarrow  \\ 
          \label{eq:5}
          &\Rightarrow& \bar{p}^G_E = {1 \over n} \sum_{i=1}^n p^G_E(i); 
        \end{eqnarray}
      with $\delta (x, y)$ being Dirac's delta. Note that $p^{G}(k; i)$ has a subtle dependency on the environment (because $G$ may react to it) and that $\bar{p}^G_E$ indicates the average probability that guesser G successfully transmits a bit through channel $E$.

      Thanks to these equations we can connect with the cost and reward functions introduced before. For every bit that attempts to be transmitted, a cost $c$ is paid. A reward $r = c/\alpha$ is cashed in only if that bit is successfully received. $\alpha$ is a parameter that controls the payoff. The survival function reads:
        \begin{eqnarray}
          \label{eq:6}
          \rho^G_E(\alpha) &=& (\bar{p}^G_E - \alpha)r, 
        \end{eqnarray}
      and $\bar{p}^G_E$ can be read from equation \ref{eq:5}. As a rule of thumb, if $\bar{p}^G_E > \alpha$ the given guesser fares well enough in the proposed environment.

      It is useful to quantify the entropy per bit of the messages produced by G: 
        \begin{eqnarray}
          \label{eq:7}
          H(G) &=& - {1 \over n}\sum_{i=1}^n \sum_k p^{G}(k; i) log\left( p^{G}(k; i) \right), 
        \end{eqnarray}
      and the mutual information between the messages and the environment: 
        \begin{eqnarray}
          \label{eq:8}
          I(G:E) &=& {1 \over n}\sum_{i=1}^n \sum_{k,k'} p_{G,E}(k, k'; i) \times \nonumber \\
          && \log \left( p_{G,E}(k, k'; i) \over p^{G}(k; i) p_E(k'; i) \right). 
        \end{eqnarray}

      To evaluate the performance of a guesser over an ensemble $\hat{E}^m$ of environments (instead  of over single environments) we attempt $N_g$ broadcasts over each of $N_e$ different environments  ($E_l \in \hat{E}^m$, $l=1, \dots, N_e \equiv ||\hat{E}^m||$) of a given size. For simplicity, instead of labeling $b^j_{i,l}$, we stack together all $N_g \times N_e$ $n$-sized words $W^j$ and $B^j$. This way $b^j_i \in B^j$ and $w^j_i \in W^j$ for $i=1, \dots, n$ and $j = 1, \dots, N_g N_e$. We have $p^G(k; i)$, $p_{\hat{E}^m}(k'; i)$, $p_{G,\hat{E}^m}(k, k'; i)$, $p^G_{\hat{E}^m}(i)$, and  $\bar{p}^G_{\hat{E}^m}$ defined just as in equations \ref{eq:2}-\ref{eq:5}, only with $j$ running through $j=1, \dots, N_g N_e$.  Also as before, we average the payoff across environments to determine whether a guesser's messages get successfully transmitted or not given $\alpha$ and the length $m$ of the environments in the ensemble:
        \begin{eqnarray}
          \label{eq:9}
          \rho^G_{\hat{E}^m}(\alpha) &=& (\bar{p}^G_{\hat{E}^m} - \alpha)r. 
        \end{eqnarray}

      Note that 
        \begin{eqnarray}
          \label{eq:20}
          I(G:\hat{E}^m) &=& {1 \over n}\sum_{i=1}^n \sum_{k,k'} p_{G,\hat{E}^m}(k, k'; i) \times \nonumber \\ 
          && log \left( p_{G,\hat{E}^m}(k, k'; i) \over p^G(k; i) p_{\hat{E}^m}(k'; i) \right) 
        \end{eqnarray}
      is different from 
        \begin{eqnarray}
          \label{eq:21}
          \left< I(G:E) \right>_{\hat{E}^m} &=& {1 \over N_e} \sum_{l=1}^{N_e} I(G:E_l). 
        \end{eqnarray}
      We use $\left< \cdot \right>_{\hat{E}^m}$ to indicate averages across environments of an ensemble $\hat{E}^m$.\\

      Finally, we discuss the generative models at the core of our bit-guessers. These are mechanisms that produce $n$-sized strings of bits, partly as a reaction to contingencies of the environment. Such message-generating processes $\Gamma^G$ could be implemented in different ways, including Artificial Neural Networks (ANNs) \cite{Hopfield1988}, spiking neurons \cite{MaassBishop2001}, Bayesian networks \cite{Pearl1985, Jensen1996}, Turing machines \cite{Turing1936}, Markovian chains \cite{Markov1971}, $\epsilon$-machines \cite{CrutchfieldYoung1989}, Random Boolean Networks (RBNs) \cite{Kauffman1993}, among others. These devices elaborate their guesses through a series of algorithms (e.g. back-propagation, message passing, or Hebbian learning) provided they have access to a sample of their environment.  

      In the real world, trial and error and evolution through natural selection would be the algorithm wiring the $\Gamma^G$ (or, in a more biological language, a genotype) into our agents. The dynamics of such evolutionary process are very interesting. However, in this paper we aim at understanding the limits imposed by a channel's complexity and the cost of inference, not the dynamics of how those limits may be reached. Therefore, we assume that our agents perform an almost perfect inference given the environment where they live. This best inference will be hard-wired in the guesser's generative model $\Gamma^G$ as explained right ahead.

      A guesser's generative model usually depends on the environment where it is deployed, so we note $\Gamma^G \equiv \Gamma^G_E$. This $\Gamma^G_E$ will consist of a pool of bits $g_{i} \in \Gamma^G_E$ (figure \ref{fig:3}) and a series of rules dictating how to emit those bits: either in a predetermined order or as a response to the channel's changing conditions. Whenever we pick up an environment  $E = \{e_i, i=1, \dots, m\}$, the best first guess possible will be the bit ($0$ or $1$) that shows up with more frequency. Hence:
        \begin{eqnarray}
          \label{eq:22}
          \Gamma^G_E(1) \equiv g_1 = \max_{k'} \left\{ p_E(k'; 1) \right\}; 
        \end{eqnarray}
      If both $0$ and $1$ appear equally often we choose $1$ without loss of generality. If the agent succeeds in its first guess, its safest next bet is to emit the bit ($0$ or $1$) that more frequently follows $g_1$ in the environment. We proceed similarly if the first two bits have been correctly guessed, if the first three bits have been correctly guessed, etc. We define $p_{B|\Gamma} (k; i)$ as the probability of finding $k = \{0,1\}$ at the $i$-th position of the $B^j$ word extracted from the environment, provided that the guess so far is correct:
        \begin{eqnarray}
          \label{eq:23}
          p_{B|\Gamma} (k'; i) &=& {1 \over Z(i)} \sum_{j=1}^m \delta(b^j_{i}, k') \prod_{i'=1}^{i-1} \delta(b^j_{i'}, g_{i'}). 
        \end{eqnarray}
      The index $j$, in this case, labels all $n$-sized words within the environment $(b^j_i \in B^j)\subset E$ and $Z(i)$ is a normalization constant that depends on how many words in the environment match $\Gamma^G_E$ up to the ($i-1$)th bit:
        \begin{eqnarray}
          \label{eq:23.1}
          Z(i) &=& \sum_{j=1}^m \prod_{i'=1}^{i-1} \delta(b^j_{i'}, g_{i'}). 
        \end{eqnarray}
      It follows:
        \begin{eqnarray}
          \label{eq:24}
          \Gamma^G_E(i=2, \dots, n) \equiv g_i &=& \max_{k'} \left\{ p_{B|\Gamma} (k'; i) \right\}. 
        \end{eqnarray}

      Note that the pool of bits in $\Gamma^G_E$ consists of an $n$-sized word, which is what they try to emit through (i.e. it constitutes the guess about) the channel. If a guesser would not be able to react to environmental conditions, the word $W$ that is actually generated at every emission would be the same in every case and $w^j_i = g_i$ always; but we allow our guessers a minimal reaction if one of the bits fails to get through (i.e. if one of the guesses is not correct). This minimal reaction capacity by our guessers results in:
        \begin{eqnarray}
          \label{eq:25}
          w^j_i &=& \Gamma^G_E(i-l) = g_{i-l}, 
        \end{eqnarray}
      where $l$ is the largest $i$ at which $w^j_i \ne b^j_i$. This means that a guesser restarts the broadcast of $\Gamma^G_E$ whenever it makes a mistake\footnote{Note that more elaborated guessers would not only reset their guess. They might browse through a tree with conditional instructions at every point. Besides an extended memory to store the growing number of branches, they would also require nested {\em if-else} instructions. On the other hand, ANNs or Bayesian networks might implement such tree-browsing without excessive {\em if-else} costs. }. 

      All together, our guesser consists of a generative model $\Gamma^G$ that contains a pool of bits and a simple conditional instruction. This is reflected in the flow chart in figure \ref{fig:3}. \\

      We have made a series of choices regarding how to implement environmental conditions. These choices affect how some randomness enters the model (reflected in the fact that, given an environment $E$, a guesser might come across different words $B^j \subset E$) and also how we implement our guessers (including their minimal adaptability to wrong guesses). We came up with a scheme that codes guessers, environments (or channels), and messages as bit strings. This allows us a direct measurement of information-theoretical features which are suitable for our discussion, but the conclusions at which we arrive should be general. Survival will depend on an agent's ability to embody {\em meaningful} information about its environment. This will ultimately be controlled by the underlying cost-efficiency tradeoff. 

      Because of the minimal implementation discussed, all bit-guessers of the same size are equal. Environmental ensembles of a given size are considered equivalent as well. Hence, the notation is not affected if we identify guessers and environments by their sizes. Accordingly, in the following we substitute the labels $G$ and $E$ by the more informative ones $n$ and $m$ respectively. Hence $\rho^G_{E_m}(\alpha)$ becomes $\rho^n_m(\alpha)$, $\bar{p}^G_E$ becomes $\bar{p}^n_m$, etc.

  \section{Results}
    \label{sec:3}

    The question that motivates this paper relates to the tradeoff between fast replication versus the cost of complex inference mechanisms. To tackle this we report a series of numerical experiments. Some of them deal with guessers in environment ensembles of fixed size, others allow guessers to switch between environment sizes to find a place where to thrive. 

    Our core finding is that the complexity of the guessers that can populate a given environment is determined by the complexity of the later. (In information theoretical terms, the complexity of the most efficiently replicated message follows from the predictability of the channel.) Back to the fast replication vs complexity question, we find environments for which simple guessers die off, but in which more complex life flourishes -- thus offering a quantifiable model for real-life excursions in biological complexity. 

    Besides verifying mathematically that the conditions for complex life exist, out model allows us to explore and quantify when and how guessers may be pushed to $m$-environments of one size or another. We expect to use this model to investigate this question in future papers. As neat examples, at the end of this paper we report i) the evolutionary dynamics established when guessers are forced to compete with each other, and ii) how the fast replication vs complexity tradeoff is altered when resources can be exhausted. These are two possible evolutionary drivers of complex life, as our numerical experiments show.

    \subsection{Numerical limits of guesser complexity}
      \label{sec:3.1}

\begin{figure}
        \begin{center}
          \includegraphics[width=0.45\textwidth]{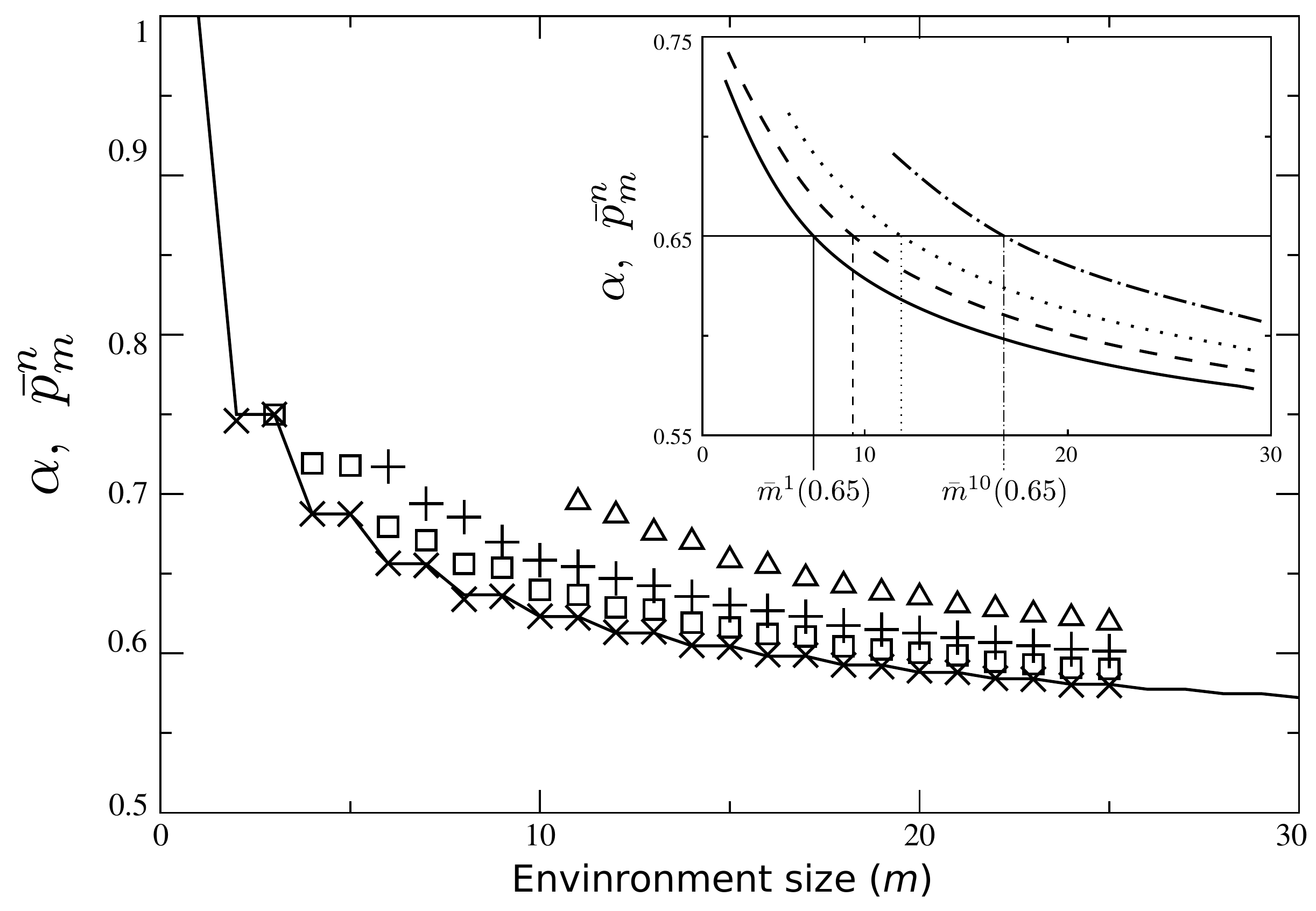}

          \caption{{\bf Probability of correctly guessing a bit in environment ensembles of constant size. } $\bar{p}^n_m$, average probability that $n$-guessers correctly guess $1$ bit in $m$-environments for different $n$ values. Here $\bar{p}^1_m$ can be computed analytically (solid line in the main plot) and marks an average, lower predictability boundary for all guessers. In the inset, the data has been smoothed and compared to a given value of $\alpha$ (represented by a horizontal line). At the intersection between this line and $\bar{p}^n_m$ we find $\bar{m}^n(\alpha)$, the environment size at which $n$-sized agents guess just enough bits to survive given $\alpha$. Notice that $n$-guessers are evaluated only in environments of size $m \ge n$. }

          \label{fig:4}
        \end{center}
      \end{figure}

      Figure \ref{fig:4} shows $\bar{p}^n_m$, the average probability that $n$-guessers correctly guess $1$ bit in $m$-environments. The $1$-guesser (that lives off maximally decorrelated bits given the environment) establishes a lower bound. More complex machines will guess more bits on average, except for infinite environment size $m \rightarrow \infty$, at which point all guessers have equivalent predictive power. 

      As $m$ grows, environments get less and less predictable. Importantly, the predictability of shorter words decays faster than that of larger ones, thus enabling guessers with larger $n$ to survive where others would perish. There are $2^n$ possible $n$-words, of which $m$ are realized in each $m$-environment. When $m >> 2^n$, the environment implements an efficient, ergodic sampling of all $n$-words -- thus making them maximally unpredictable. When $n \lesssim m < 2^n$ the sampling of $n$-sized words is far from ergodic and a non-trivial structure is induced in the environment because the symmetry between $n$-sized words is broken -- they cannot be equally represented due to finite size sampling effects.
  
      This allows that complex guessers (those with the ability to contemplate larger words, keep them in memory, and make choices regarding information encoded in larger strings) can guess more bits, on average, than simpler agents. In terms of messages crossing the channel, while shorter words are meaningless and basically get transmitted (i.e. are correctly guessed) by chance alone, larger words might contain meaningful, non-trivial information that get successfully transmitted because they cope with the environment in an adequate way.

      Note that this symmetry breaking to favor predictability of larger words is just a mechanism that allows us to introduce correlations in a controlled and measurable way. In the real world this mechanism might correspond to asymmetries between dynamical systems in temporal or spatial scales. Although our implementation is rather ad hoc (suitable to our computational and conceptual needs), we propose that similar mechanisms might play important roles in shaping life and endowing the universe with meaningful information. Indeed, it might be extremely rare to find a kind of environment in which words of all sizes become non-informative simultaneously.\\

\begin{figure}
        \begin{center}
          \includegraphics[width = 0.5\textwidth]{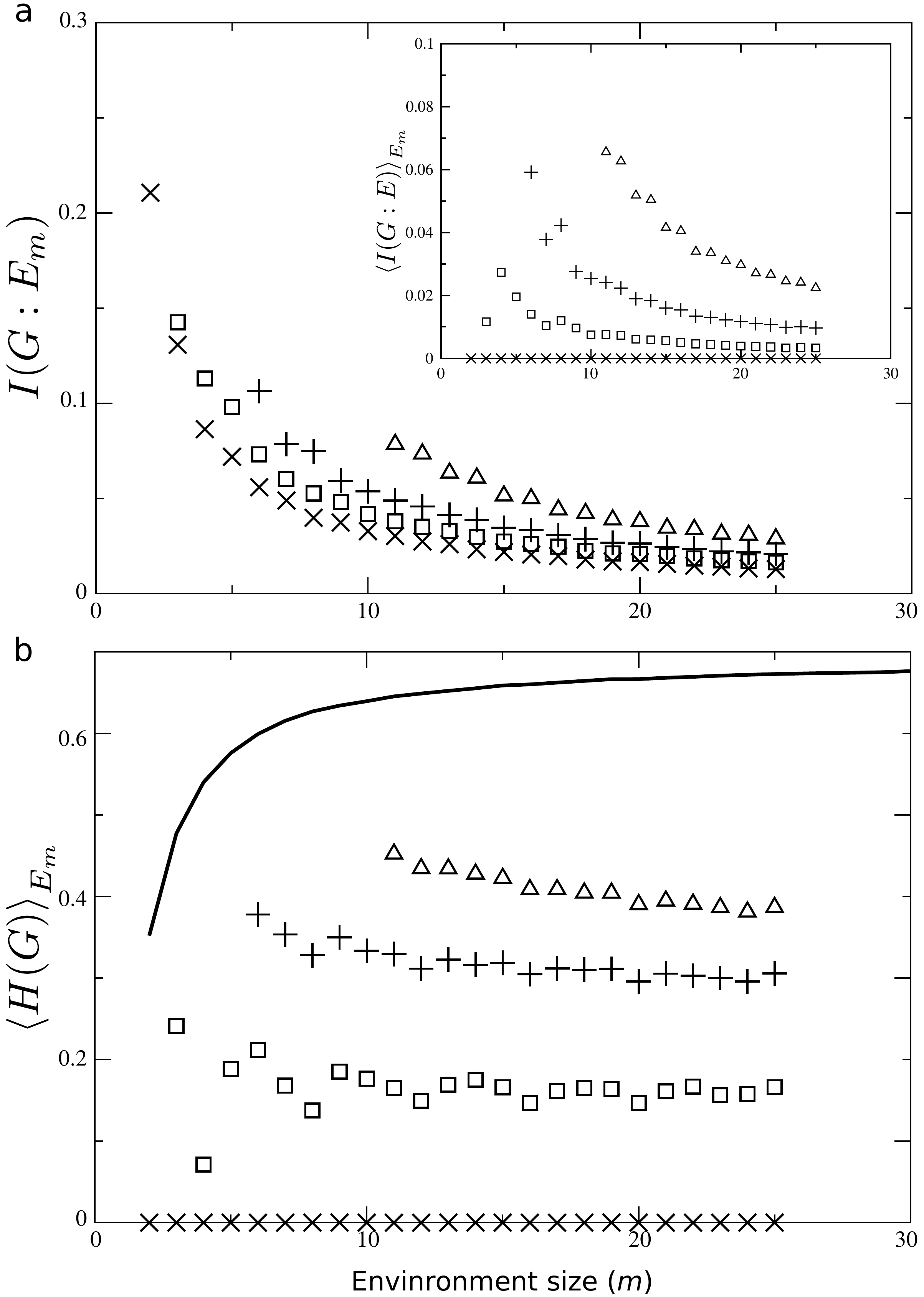}

          \caption{{\bf Mutual information and entropy. } Guessers with $n=1$ (crosses), $n=2$ (squares), $n=5$ (pluses), and $n=10$ (triangles) are presented. {\bf a} $I(G:E_m)$ and $\left< I(G:E) \right>_{E_m}$ (inset) quantify the different sources of information that allow more complex guessers to thrive in environments in which simpler life is not possible. {\bf b} The entropy of a guesser's message given its environment seems roughly constant in these experiments despite the growing environment size. This suggests an intrinsic measure of complexity for guessers. Larger guessers look more random even if they might carry more meaningful information about their environment. The thick black line represents the average entropy of the environments (which approaches $log(2)$) against which the entropy of the guessers can be compared. }

          \label{fig:5}
        \end{center}
      \end{figure}

      The mutual information between a guesser's response and the environment (i.e. between broadcast messages and channel conditions) further characterizes the advantages of more complex replicators. Figure \ref{fig:5}{\bf a} shows $I(G:E_m)$ and $\left< I(G:E) \right>_{E_m}$. As we noted above, these quantities are not the same. Let us focus on $1$-guessers for a moment to clarify what these quantities encode. 

      Given an $m$-environment, $1$-guessers have got just one bit that they try to emit repeatedly. They do not react to the environment -- there is not room for any reaction within one bit, so their guess is persistently the same. The mutual information between the emitted bit and the arbitrary words $B \subset E$ that $1$-guessers come across is precisely zero, as shown in the inset of figure \ref{fig:5}{\bf a}. Hence, $\left< I(G:E) \right>_{E_m}$ captures the mutual information due to the slight reaction capabilities of guessers to the environmental conditions. 

      While the bits emitted by $1$-guessers do not correlate with $B \subset E$, they do correlate with each given $E$ since they represent the most frequent bit in the environment. Accordingly, the mutual information between a $1$-guesser and the aggregated environments (reflected by $I(G:E_m)$) is different from zero (figure \ref{fig:5}{\bf a}). To this quantity contribute both the reaction capability of guessers and the fact that they have hard-wired a near-optimal guess in $\Gamma^G_E$, as explained in section \ref{sec:2.1}.

      We take the size of a guesser $n$ as a crude characterization of its complexity. This is justified because larger guessers can store more complex patterns. $\left< H(G)\right>_{E_m}$ indicates that more complex guessers look more entropic than less complex ones (figure \ref{fig:5}{\bf b}). Larger guessers come closer to the entropy level of the environment (black thick line in figure \ref{fig:5}{\bf b}), which itself tends rapidly to $log(2)$ per bit. Better performing guessers appear more disordered to an external observer even if they are better predictors when considered within their context. Note that $\left< H(G)\right>_{E_m}$ is built based on the bits actually emitted by the guessers. In biological terms, this would mean that this quantity correlates with the complexity of the phenotype. For guessers of fixed size $n$, we observe a slight decay of $\left< H(G)\right>_{E_m}$ as we proceed to larger environments. \\

      The key question is whether the payoff may be favorable for more complex guessers provided that they need a more costly machinery in order to get successfully reproduced. As discussed above, if we would use, e.g., Artificial Neural Networks or Bayesian Inference Graphs to model our guessers, a cost could be introduced for the number of units, nodes, or hidden variables. These questions might be worth studying somewhere else. Here we are interested in the mathematical existence of such favorable tradeoff for more complex life. To keep the discussion simple bit guessers incur only in a cost proportional to the number of bits that they try to transmit. Note that we do not lose generality because such limit cost shall always exist. Equation \ref{eq:9} captures all the forces involved: the cost of transmitting longer messages versus the reward of a successful transmission. 

      Guessers of a given size survive in an environment ensemble if, on average, they can guess enough bits of the environment or, using the information theory picture, if they can convey enough bits through the channel (in any case, they survive if $\bar{p}^n_m > \alpha$, which implies $\rho^n_m>0$). Setting fix a value of $\alpha$ we find out graphically $\bar{m}^n(\alpha)$, the largest environment at which $n$-guessers survive (figure \ref{fig:4}, inset). Because $m$-environments look more predictable to more complex guessers we have that $\bar{m}^n(\alpha) > \bar{m}^{n'}(\alpha)$ if $n>n'$. This guarantees that for $\alpha > 0.5$ there always exist $m$-environments from which simple life is banned while more complex life can thrive -- i.e. situations in which environmental complexity is an explicit driver towards more complex life forms. 

      This is the result that we sought. The current model allows us to illustrate mathematically that limit conditions exist under which more complex and costly inference abilities can overcome the pressure for fast and cheaper replication. Also, the model allows for explicit, information theoretically-based quantification of such a limit.

    \subsection{Evolutionary drivers}
      \label{sec:3.2}

      Despite its laborious mathematical formulation, we think that our bit-guesser model is very simple and versatile. We think that it can easily capture fundamental information-theoretical aspects of biological systems. In future papers we intend to use it to further explore relationships between guessers and environments, within ecological communities, or in more simple symbiotic or parasitic situations. To illustrate how this could work out we present now some minimal examples. \\

      Let us first explore some dynamics in which guessers are encouraged to explore more complex environments, but this same complexity can become a burden. As before, let us evaluate an $n$-guesser $N_g \cdot N_e$ times in a sample of the $m$-environment ensemble. Let us also look at $\hat{\rho}^n_m(\alpha, N_g, N_e)$, the accumulated reward after these $N_g \cdot N_e$ evaluations -- note that $\hat{\rho}^n_m$ is an empirical random variable now. If $\hat{\rho}^n_m(\alpha, N_g, N_e) > 0$, the $n$-guesser fares well enough in this $m$-environment and it is encouraged to explore a more complex one. As a consequence, the guesser is promoted to an $(m+1)$-environment, where it is evaluated again. If $\hat{\rho}^n_m(\alpha, N_g, N_e) < 0$, this $m$-environment is excessively challenging for this $n$-guesser, and it is demoted to an $(m-1)$-environment. Note that the $n$-guesser itself remains with a fixed size throughout. It is the complexity of the environment that changes depending on the reward accumulated. 

\begin{figure}
        \begin{center}
          \includegraphics[width = 0.5\textwidth]{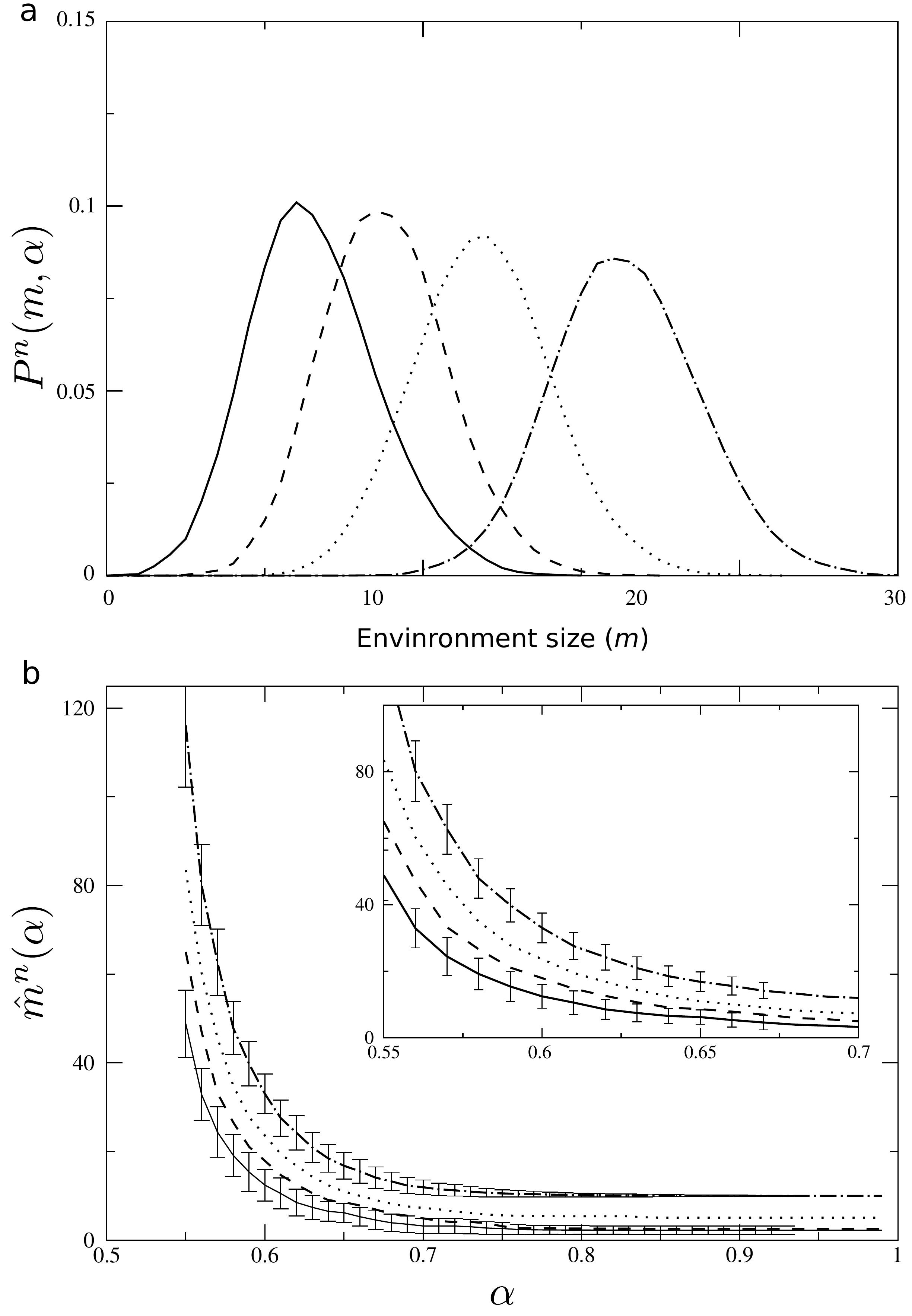}

          \caption{{\bf Dynamics around $\bar{m}^n(\alpha)$. } Again, guessers with $n=1$ (solid line), $n=2$ (dashed line), $n=5$ (dotted line), and $n=10$ (dot-dashed line). {\bf a} $P^n(m, \alpha)$ tells us how often do we find $n$-guessers in $m$-environments when they are allowed to roam constrained only by their survival function $\rho^n_m$. The central value $\hat{m}^n$ of $P^n(m, \alpha)$ must converge to $\bar{m}^n(\alpha)$ and oscillations around it depend (through $N_g$ and $N_e$) on how often do we evaluate the guessers in each environment. {\bf b} Average $\hat{m}^n$ for $n=1,2,5,10$ and standard deviation of $P^n(m, \alpha)$ for $n=1, 10$. Deviations are not presented for $n=2, 5$ for clarity. The inset represents a zoom in into the main plot. }

          \label{fig:6}
        \end{center}
      \end{figure}

      As we repeatedly evaluate the $n$-guesser, some dynamics are established which let the guesser explore more or less complex environments. The steady state of these dynamics is characterized by a distribution $P^n(m, \alpha)$. This tells us the frequency with which $n$-guessers are found in environments of a given size (figure \ref{fig:6}{\bf a}). Each $n$-guesser has its own distribution that captures the environmental complexity that the guesser deals more comfortably with. The overlaps and gaps between $P^n(m, \alpha)$ for different $n$ suggest that: i) some guessers would engage in harsh competition if they needed to share environments of a given kind and ii) there is room for different guessers to get segregated into environments of increasing complexity. 

      The average
        \begin{eqnarray}
          \label{eq:26}
          \hat{m}^n(\alpha) &=& \sum_{m} m P^n(m, \alpha) 
        \end{eqnarray}
      should converge to $\hat{m}^n(\alpha) \simeq \bar{m}^n(\alpha)$ under the appropriate limit. This is, if we evaluate the guessers numerically enough times, the empirical value $\hat{m}^n(\alpha)$ should converge to the mean field value $\bar{m}^n(\alpha)$ shown in the inset of figure \ref{fig:4}. Figure \ref{fig:6}{\bf b} shows dynamically-derived averages $\hat{m}^n(\alpha)$ and some deviations around them as a function of $\alpha$.\\

      It is easily justified that guessers drop to simpler environments if they cannot cope with a large complexity. It is less clear why they should seek more complicated environments if they thrive in a given one. This might happen if some external force drives them. For example if simpler guessers (which might be more efficient in simpler environments) have already crowded the place. Let us remind, from figure \ref{fig:4}, how given an environment size more complex guessers can always accumulate a larger reward. This might suggest that complex guessers always pay off, but the additional complexity might become a burden in energetic terms -- consider, e.g., the exaggerated metabolic cost of mammal brains. It is non-trivial how competition dynamics between guessers of different size can play out. Let us gain some insights by looking at a simple model. 

      $n$-guessers with $n = 0$, $1$, $2$, $3$, and $4$ were randomly distributed occupying $100$ environments, all of them with fixed size $m$. These guessers were assigned an initial $\hat{\rho}_i(t=0) = n\rho_0$. Here, $i=1, \dots, 100$ labels each one of the $100$ available guessers. Larger guessers start out with larger $\hat{\rho}_i(t=0)$ representing that they come into being with a larger metabolic load satisfied. A $0$-guesser represents an unoccupied environment. New empty environments might appear only if actual ($n \ne 0$) guessers die, as we explain below. We tracked the population using $P^m(n, t)$, the proportion of $0$-, $1$-, $2$-, $3$-, and $4$-guessers through time\footnote{These experiments were the more computationally demanding, that is why we took $n=1, 2, 3, 4$ instead of the values $n=1, 2, 5, 10$ used throughout the paper. The insights gained from the simulations do not depend on the actual values of $n$.}. 

      \begin{figure*}
        \begin{center}
          \includegraphics[width = \textwidth]{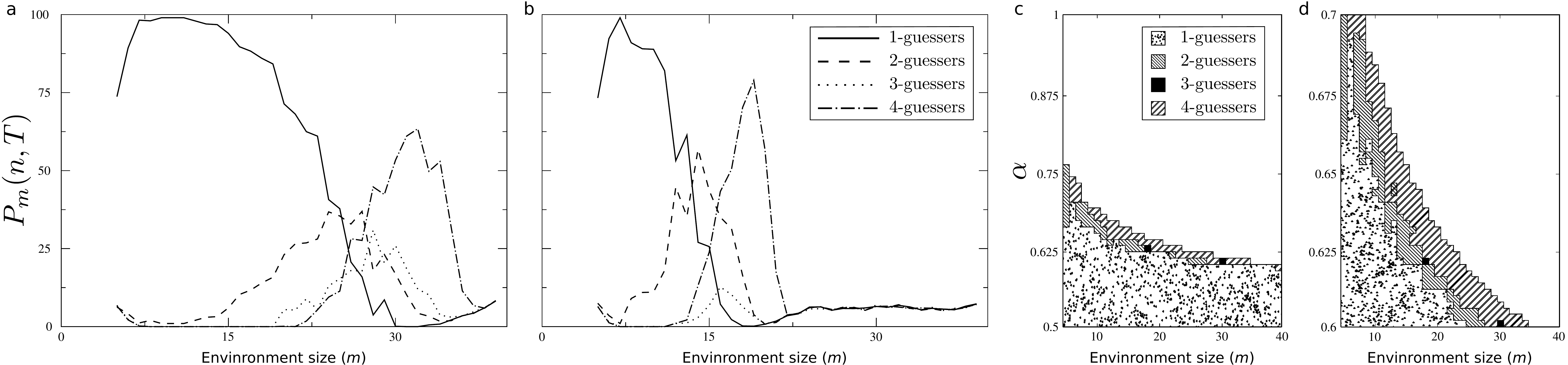}

          \caption{{\bf Evolutionary drivers: competition. } Coexisting replicators will affect each other's environments in non-trivial ways which may often result in competition. We implement a dynamics in which $1$-, $2$-, $3$-, and $4$-guessers exclusively occupy a finite number of environments of a given size (fixed $m$). The $100$ available slots are randomly occupied at $t=0$ and granted to the best replicators as the dynamics proceed. We show $P_m(n, t=10\>000)$ for $m = 5, \dots, 39$ and $\alpha = 0.6$ ({\bf a}), $\alpha = 0.65$ ({\bf b}). The most abundant guesser at $t=10\>000$ is shown for $\alpha \in (0.5, 1)$ ({\bf c}) and $\alpha \in (0.6, 0.7)$ ({\bf d}). Once $m$ is fixed, there is an upper value of $\alpha$ above which no guesser survives and all $100$ available slots remain empty. Competition and the replication-predictability tradeoff segregate guessers according to the complexity of the environment -- i.e. of the transmission channel. Coexistence of different guessers seems possible (e.g. $m=15$ in {\bf b}), but it cannot be guaranteed that the dynamics have converged to a steady distribution. }

          \label{fig:7}
        \end{center}
      \end{figure*}

      At each iteration, a guesser (say the $i$-th one) was chosen randomly and evaluated with respect to its environment. Then the wasted environment was replaced by a new, random one with the same size. We ensured that every guesser attempts to guess the same amount of bits on average. This means, e.g., that $1$-guessers are tested twice as often as $2$-guessers, etc. If after the evaluation we found that $\hat{\rho}_i(t+\Delta t) < 0$, then the guesser died and it was substituted by a new one. The $n$ of the new guesser was chosen randomly after the current distribution $P^m(n, t)$. If $\hat{\rho}_i(t+\Delta t) > 2n\rho_0$, the guesser got replicated and shared its $\hat{\rho}_i$ with its daughter, who overrode another randomly chosen guesser. This replication at $2n\rho_0$ represents that, before creating a similar agent, parents must satisfy a metabolic load that grows with their size. There is a range ($0 < \hat{\rho}_i < 2n\rho_0$) within which guessers are alive but do not replicate. 

      Of course, this minimal model is just a proxy and softer constraints could be placed. These could allow, e.g., for random replication depending on the accumulated $\hat{\rho}_i(t+\Delta t)$, or for larger progeny if $\hat{\rho}_i(t+\Delta t) >> 2n\rho_0$. These are interesting variations that might be worth exploring. There are also some insights to be gained from the simple setup considered here. We expect that more complex models will largely inherit the exploratory results that follow. 

      Figure \ref{fig:7}{\bf a} and {\bf b} show $P_m(n, t=10\>000)$ with $\alpha = 0.6$ and $0.65$. Note that for large environments all guessers combined do not add up to $100$. Indeed, they fall short from that number -- i.e. mostly empty slots remain. The most abundant guesser after $10\>000$ iterations is shown in figure \ref{fig:7}{\bf c} as a function of $m$ and $\alpha$. 

      These plots show how guessers naturally segregate in environments depending on their complexity, with simpler guessers crowding simpler environments as suggested above. In such simple environments, the extra reward earned by more complex guessers does not suffice to overcome their energetic cost and they lost in this direct competition. They are, hence, pushed to more complex environments where their costly inference machinery pays off. 

      After $10\>000$ iterations we also observe cases in which different guessers coexist. This means that the mathematical limits imposed by this naive model do not imply an immediate, absolute dominance of the fittest guesser. Interesting temporal dynamics might arise and offer the possibility to model complex ecological interactions. \\

      So far our guessers only interacted with the environment in a passive way, by receiving the reward that the corresponding $m$-environment dictates. But living systems also shape their niche in return. Such interplay can become very complicated and we think that our model offers a powerful exploratory tool. Let us study a very simple case in which the actions of the guessers (i.e. their correctly guessing a bit or not) affect the reward that an environment can offer. 

\begin{figure}
        \begin{center}
          \includegraphics[width = 0.5\textwidth]{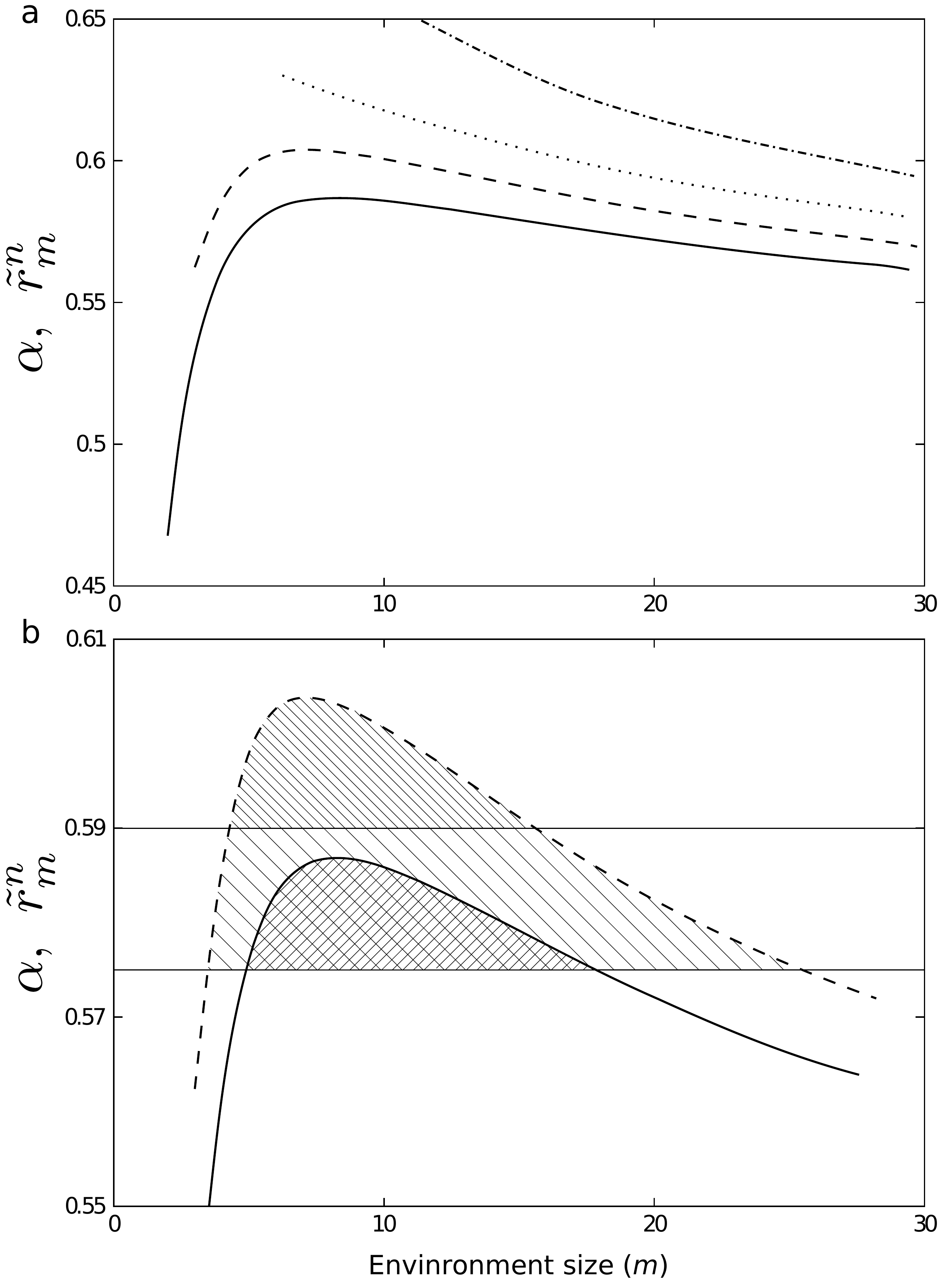}

          \caption{{\bf Evolutionary drivers: exhausted resources.} Rather than monopolizing channel slots (as in figure \ref{fig:5}), we can also conceive individual bits as valuable, finite resources that get exhausted whenever they are correctly {\em guessed}. Then a successful replicator can spoil its own environment and new conditions might apply to where life is possible. {\bf a} Average reward obtained by $1$-, $2$-, $5$-, and $10$-guessers in environments of different sizes when bits get exhausted with efficiency $\beta = 1$ whenever they are correctly guessed. {\bf b} Given $\alpha = 0.575$ and $\alpha = 0.59$, $1$- and $2$-guessers can survive within upper and lower environment sizes. If the environment is too small, resources get consumed quickly and cannot sustain the replicators. In message transmission language, the guessers crowd their own channel. If the environment is too large, unpredictability takes over for these simple replicators and they perish.}

          \label{fig:8}
        \end{center}
      \end{figure}

      To do so we rethink the bits in an environment as resources that can be exhausted if correctly guessed, but also replenished after enough time has elapsed. Alternatively, thinking from the message broadcasting perspective, a spot on the channel might appear crowded if it is engaged in a successful transmission. Assume that every time that a bit is correctly guessed it gets exhausted (or gets crowded) with an efficiency $\beta$ so that on average each bit cannot contribute any reward $\beta (\bar{p}^n_m/m)$ of the time. The average reward extracted by a guesser from an ensemble becomes:
        \begin{eqnarray}
          \label{eq:27}
          \tilde{r}^n_m &=& \left( 1 - \beta {\bar{p}^n_m \over m}  \right)\bar{p}^n_m r, 
        \end{eqnarray}
      which is plotted for $1$-, $2$-, $5$-, and $10$-guessers and $\beta = 1$ in figure \ref{fig:8}. 

      Smaller guessers living in very small environments quickly crowd their channels (alternatively, exhaust the resources they depend on). In figure \ref{fig:8}{\bf b} (still with $\beta = 1$) given some $\alpha$, $1$- and $2$-guessers can only survive within some under and upper limits (figure \ref{fig:8}{\bf b}). Furthermore, the slope of the curves around these limits also tell us important information. If these guessers dwell in environments around the lower limit (i.e. near the smallest $m$-environment where they can persist), then moving to larger environments will always report larger rewards. But if they dwell close to the upper limit, moving to larger environments will always be detrimental. In other words, dynamics such as the one introduced at the beginning of this section (illustrated in figure \ref{fig:6}{\bf a}) would have respectively unstable and stable fixed points in the upper and lower limits of persistence. 

      This simple model illustrates how scarcity of resources (and, more general, other kinds of guesser-environment interactions) might play an important role as evolutionary drivers towards more complex life. This does not intend to be an exhaustive nor a definitive model, just an illustration of the versatility of the bit-guessers and environments introduced in this paper.

  \section{Discussion}
    \label{sec:4}

    In this paper we have considered a fundamental question related to the emergence of complexity in living systems. The problem being addressed here is whether the mathematical conditions exist such that mode complex organisms can overcome the cost of their complexity by developing a higher potential to predict the external environment. As suggested by several authors \cite{Jacob1998, Wagensberg2000, Friston2013} the behavioral plasticity provided by the exploratory behavior of living systems can be understood it terms of their potential for dealing with environmental information \cite{Gerhart1997}. 

    Our models make an explicit approach by considering a replication-predictability tradeoff under very general assumptions, namely: i) More complex environments look more unpredictable to simpler replicators and ii) Agents that can keep a larger memory and make inferences based on more elaborated information can extract enough valuable bits from the environment as to survive in those more challenging situations. Despite the inevitable cost inherent to the  cognitive machinery, a selection process towards more complex life is shown to exist. This paves the way for explicit evolutionary pressures towards more complex life forms.

    In our study we identify a transmitter (replicators at a given generation), a receiver (replicators at the next generation), and a channel ({\em any} environmental conditions) through which a message (ideally instructions about how to build newer replicators) is passed on. Darwinian evolution follows naturally as effective replicators transit a channel faster and more reliably thus getting more and more space in successive generations. The inference task is implicit as the environment itself codes for meaningful bits of information that, if picked up by the replicators, boost the fitness of the phenotypes embodied by the successful messages. 

    This view is directly inspired by a qualitative earlier picture introduced by \citet{MaynardSmith2000}. That metaphor assigned to the DNA some external meaning that had to be preserved {\em against} the environmental noise. Contrary to this, we propose that, as messages attempt to travel from a generation to the next one, all channel conditions (including noise) pump relevant bits into the transmitted strings -- hence there is no need to protect meaning against the channel because, indeed, meaningful information emerges out of the replicator's interaction with such channel contingencies. The way that we introduce correlations in our scheme (through a symmetry breaking between the information borne by short and larger words due to finite size effects) is compatible with this view. However, interestingly, it also suggests that meaningful information might arise naturally even in highly unstructured environments when different spatial and temporal scales play a relevant role.

    This way of integrating information theory and Darwiniand evolution is convenient to analyze the questions at hand that concern the emergence of complex life forms. But it also suggests further research lines. As discussed at  the beginning of the paper, guessers and their transmissible messages might and should shape the transmission channel (e.g., by crowding it, as explored briefly in section \ref{sec:3.2}). What possible co-evolutionary dynamics between guessers and channels can be established? Are there stable ones, others leading to extinction, etc? Do some of them, perhaps, imply open-ended evolution? Which ones? These are questions that relate tightly to the phenomenon of niche construction. We propose that they can be easily modeled within the proposed bit-guesser paradigm. Further exploring the versatility of the model, a guesser's transmitted message might be considered an environment in itself; thus opening the door to ecosystem modeling based on bare information theory. It is also suggested the exploration of different symbiotic relationships from this perspective and how they might affect coevolution. 

    Finally, an important question was left aside that concerns the memory vs adaptability tradeoff of bit guessers. Here we studied guessers with a minimal adaptability to focus on the emerging hierarchy of complexity. Adaptability at faster (say, at behavioral) temporal scales is linked to more complex inferences with richer dynamics. This brings in new dilemmas as to how to weight the different building blocks of complex inference -- e.g. how do we compare memory and {\em if-else} or {\em while} instructions? These and other questions are left for exploration in future research.

\vspace{0.2 cm}

  \section*{Acknowledgments}

    The authors thank the members of the Complex Systems Lab, and Max Tegmark, Jeremy Owen, Henry Lin, Jordan Horowitz, and David Wolpert for very useful discussions. This study was supported by an European Research Council Advanced Grant (SYNCOM), the Botin Foundation, by Banco Santander through its Santander Universities Global Division and by the Santa Fe Institute. This work has also been supported by the Secretaria d'Universitats i Recerca del Departament d'Economia i Coneixement de la Generalitat de Catalunya.




\end{document}